\documentclass{edm_template}
\usepackage{tabu}
\usepackage[letterpaper]{geometry}

\begin{document}

\title{Modelling End-of-Session Actions in Educational Systems}
\numberofauthors{1} 
\author{
\alignauthor
Christian Hansen, Casper Hansen, Stephen Alstrup, Christina Lioma  \\ 
       \affaddr{University of Copenhagen}\\
       \affaddr{Department of Computer Science, Denmark}\\
       \email{\{chrh,c.hansen,s.alstrup,c.lioma\}@di.ku.dk}
}
\maketitle
\begin{abstract}
In this paper we consider the problem of modelling when students end their session in an online mathematics educational system. Being able to model this accurately will help us optimize the way content is presented and consumed. This is done by modelling the probability of an action being the last in a session, which we denote as the End-of-Session probability. We use log data from a system where students can learn mathematics through various kinds of learning materials, as well as multiple types of exercises, such that a student session can consist of many different activities. 
We model the End-of-Session probability by a deep recurrent neural network in order to utilize the long term temporal aspect, which we experimentally show is central for this task. Using a large scale dataset of more than 70 million student actions, we obtain an AUC of 0.81 on an unseen collection of students. Through a detailed error analysis, we observe that our model is robust across different session structures and across varying session lengths. 
\end{abstract}
%
\section{Introduction and related work}
Digitization of education is ever increasing, where for higher level education, massively open online course (MOOC) platforms are offering an increasing amount of high quality courses for a wide range of topics. Similarly, a wide variety of systems exist for assisting teachers in lower levels of education, where especially mathematics has been a focus since many types of exercises allow automatic correction, thus freeing up teacher resources. The design of most of these systems is that students engage with them through sessions consisting of different actions, e.g. answering questions or reading learning material, but the frequency and length of sessions naturally vary between students.

In this paper we consider the problem of modelling when an action is the last action in a session, which we denote as the End-of-Session action. This is naturally a highly class imbalanced problem, since there is only one End-of-Session action per session, and we do not get any explicit feedback about the student's intent to end a session before it is over. Additionally, a large amount of external and internal factors for ending a session exists, for example: The student has finished an assigned task (e.g. homework), the student is unmotivated or distracted, the student has grasped the material, etc.
If we are able to model the End-of-Session probability for each action accurately, then it can offer real-time insights in ongoing student sessions, and actions can be taken to steer the student in a direction where she is less likely to quit.

The problem of modelling when a session ends has, to the best of our knowledge, only recently been considered by Kassak et al. \cite{kassak2016student} in the educational domain, where a polynomial classification model with handcrafted features was used, but the work was done on a small scale using just 452,000 student actions. Related work to this problem has also been considered from the point of view of modelling and clustering student sessions, in order to understand and group student behaviour in educational systems \cite{shenclustering, hansensequence, klingler2016temporally}.
Even though the End-of-Session problem has not been investigated much in the educational domain, the problem has been considered in other domains, such as business \cite{tax2017predictive} and media streaming \cite{vasiloudis2017predicting,hansenSpotifyComp}, where recurrent neural networks and gradient boosted trees have been shown to work well.

A related problem to End-of-Session modelling that has been considered extensively in the educational domain, is drop-out prediction. In this setting student log data is used for modelling drop-out of students, with studies focusing on whether a student will drop out of their studies \cite{bayer2012predicting}, or drop out of MOOC courses \cite{halawa2014dropout, liang2016machine, xing2016temporal}. Drop-out prediction models can utilize both general characteristics of a student, as well as usage behaviour changes (e.g. usage decline) to aid in the prediction.
Although End-of-Session modelling utilizes much of the same types of data, in this setting the "stop"-signal needs to be found within each session, and not as a single terminal event happening at a single point.

In this paper we model the End-of-Session problem using a deep recurrent neural network architecture, that outputs an End-of-Session probability for each student action, using the actions in the current and previously completed sessions. We show that considering the task on a studentwise level obtains significant improvements over considering the sessions individually.
We experimentally evaluate our approach using log data from more than 70 million student actions from 85,780 students, where we are able to obtain an AUC score of 0.81 on an unseen collection of students. Through a detailed error analysis we show that the model is robust across students with a varying number of completed sessions, across sessions of different lengths, as well as for different session structures. 

In the remaining of the paper the data is presented in Section \ref{sec:data}; the End-of-Session problem is presented in Section \ref{sec:modeleos}, and our deep learning approach presented in Section \ref{our-approach}; experimental findings are presented and analyzed in Section \ref{sec:experimentalpart}; and a conclusion is made in Section \ref{sec:conclusion}.
\section{Data}
\label{sec:data}
In this paper we use data provided by Edulab\footnote{The data is proprietary and not publicly available.}, the largest Danish system for online mathematics. The system primarily targets students aged 6 to 16 and supplies material spanning the entire mathematics curriculum. The system offers two primary ways of engaging with the material: 1) Reading text material or watching video material, and 2) answer fill-out or multiple-choice questions.
These two broad categories can be done by students on their own initiative or by being assigned it as homework by a teacher. All questions and video/text materials are associated with a specific \emph{lesson}, which is a specific skill such as "addition of small integers", and each lesson is associated with a \emph{topic}, which is a broad skill such as "addition". The general data statistics can be seen in Table \ref{tab:dataStats}.

We use log data generated when students use the system. For each action they perform we have the timestamp, the lesson id, the topic id, the type of action (and an answer if the action is answering a question), and lastly whether it is homework. The system does not track the exact time taken to complete an action, but can derive it as the time taken since the last action. To group the actions into sessions, we consider a time gap of more than 15 minutes to be a new session. The choice of 15 minutes was based on each action rarely requiring more than a few minutes, and to allow time for breaks within the same session. From these we derive a set of general features associated with each action as described in Table \ref{tab:datasetFeatures}. For the temporal features we chose to discretize the time into the time intervals 8-12, 12-15, and 15-8 in order to represent the times as "before noon", "afternoon", and "after school". For content type we decided to ignore the actual ids, and instead focus on the changing between lessons and topics. The reason for this was the large number of different ids, with close to 1000 different lesson ids. This also has the benefit of being more general, since the system will not be able to learn patterns specific to certain ids. 

\begin{figure*}
    \centering
    \begin{minipage}{0.5\textwidth}
        \centering
        \includegraphics[width=0.8\textwidth]{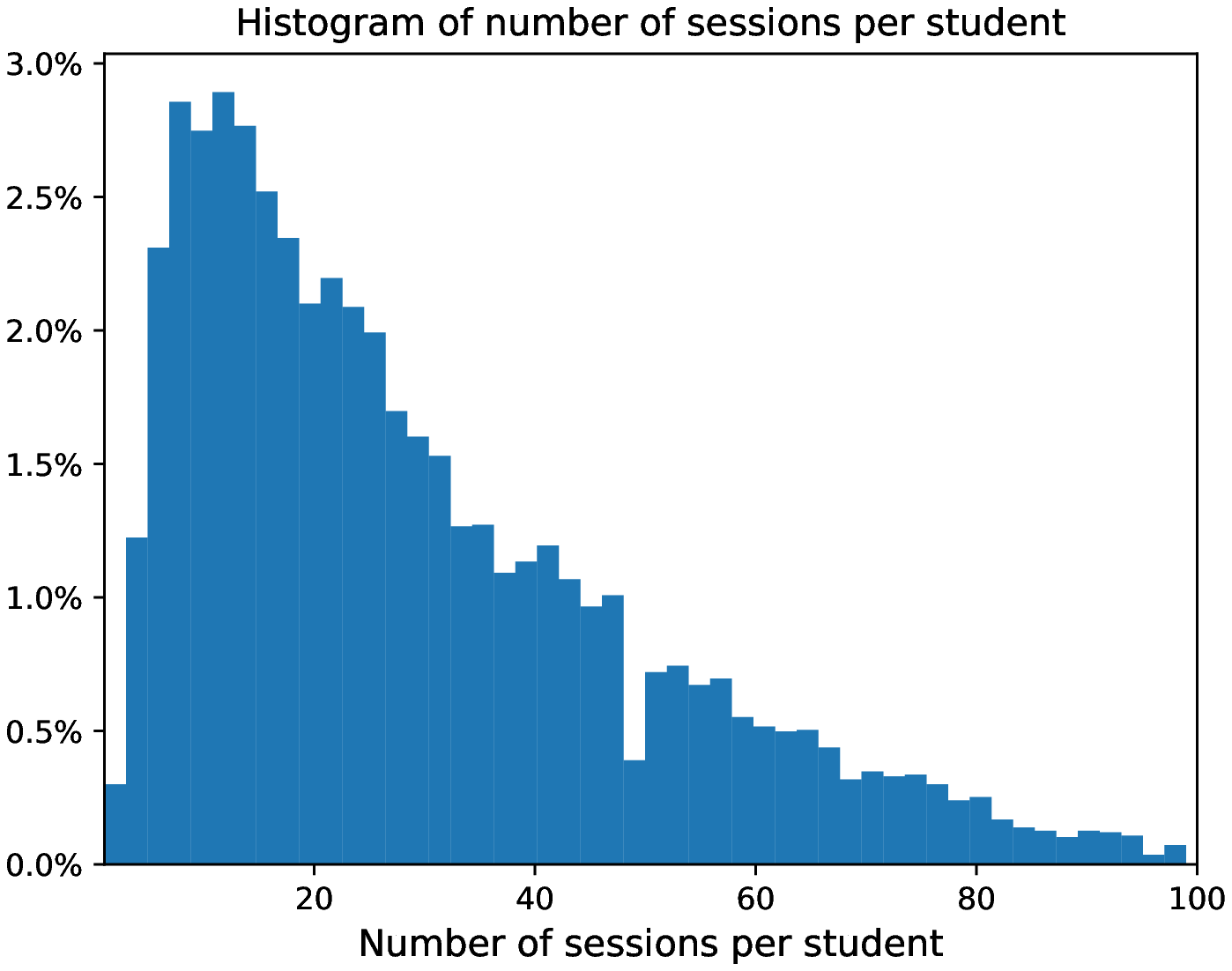} 
        \vspace{-7pt}
        \caption{The distribution of sessions per student.}
        \label{fig:numsesstudenthist}
        \vspace{-7pt}
    \end{minipage}\hfill
    \begin{minipage}{0.5\textwidth}
        \centering
        \includegraphics[width=0.8\textwidth]{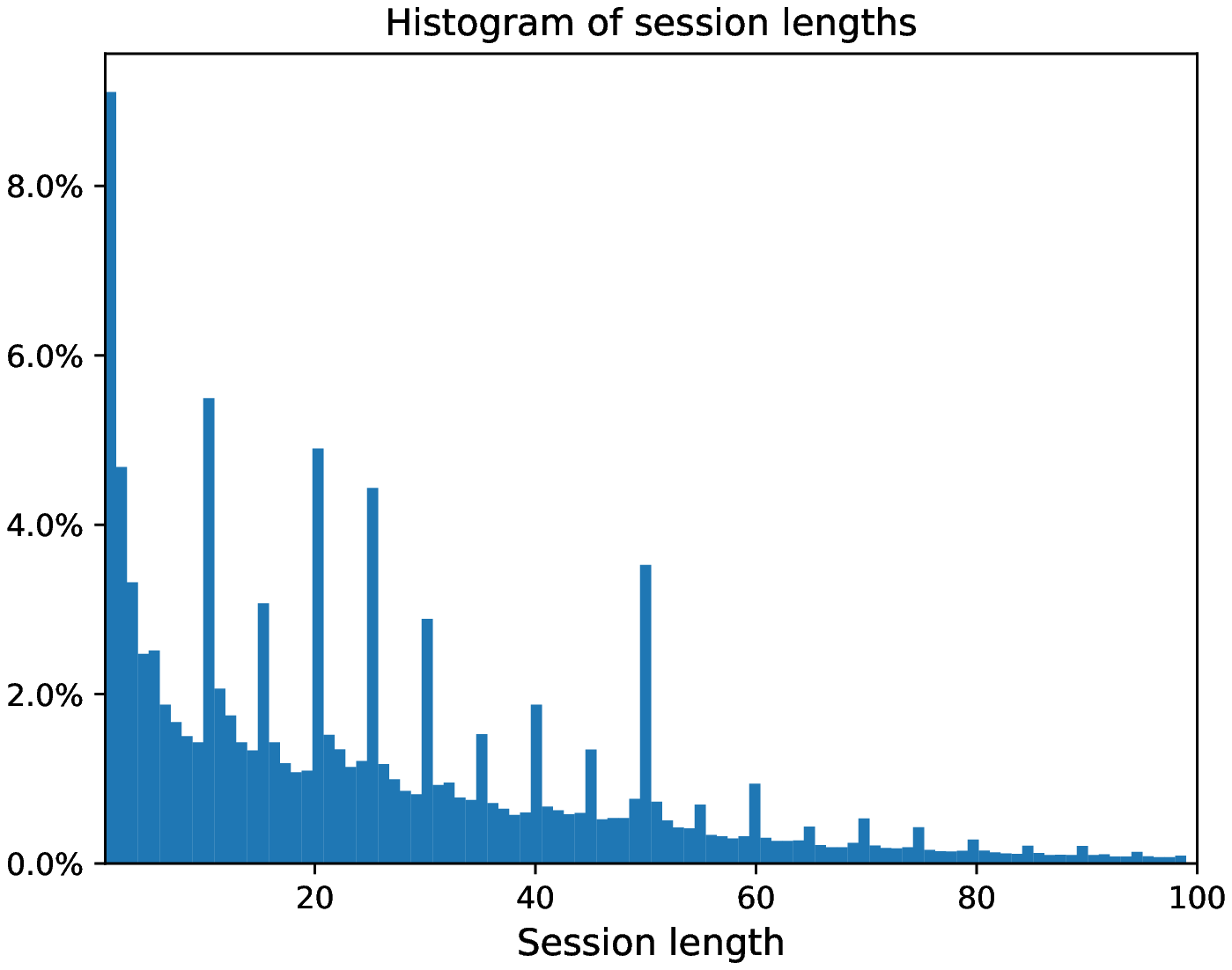} 
        \vspace{-7pt}
        \caption{The distribution of session lengths.}
        \label{fig:seslenhist}
        \vspace{-7pt}
    \end{minipage}
\end{figure*}

\begin{table}
    \centering
    \begin{tabu}{|l|X|}
    \hline
        Category & Features \\ \hline
        Temporal & 1) time of day as discrete values 8-12, 12-15, and 15-8, 2) time since last action, and 3) time since last session. \\ \hline
        
        Action type & 1) answering fill-out question, 2) answering multiple-choice question, and 3) watching or reading material. \\ \hline
        
        Content type & 1) the lesson id was changed compared to the previous action, and 2) the topic id was changed compared to the previous action. \\ \hline
        
        Miscellaneous & 1) whether a potential question is answered correctly, and 2) if the action was done as homework or on the student's own initiative. \\ \hline
    \end{tabu}
    \caption{Dataset features divided into categories.}
    \label{tab:datasetFeatures}
\end{table}
There is a large variance in the number of sessions each student has completed and the length of each of the sessions, the histograms for each of these can be seen in Figure \ref{fig:numsesstudenthist} and \ref{fig:seslenhist} respectively. For the histogram of session lengths, spikes are seen at certain lengths divisible by 5, this is due to how some groups of questions are presented to the students, e.g. as 5 or 10 quick related questions. Similarly, homework often consists of a number of questions divisible by 5.
\begin{table}
    \centering
    \scalebox{0.8}{
    \begin{tabular}{|l|l|}
    \hline
        Number of students & 85,780 \\ \hline
        Number of sessions & 2,587,876 \\ \hline
        Number of actions & 71,341,770 \\ \hline
        \hline
        Percentage of sessions being homework & 48.3\% \\ \hline
        Percentage of sessions being partly homework & 25.5\% \\ \hline
        Percentage of sessions not being homework & 26.2\% \\ \hline
    \end{tabular}}
    \caption{General data statistics.}
    \label{tab:dataStats}
    \vspace{-10pt}
\end{table}
\section{Modelling End-of-Session}
\label{sec:modeleos}
In this section we present the problem of modelling when a session ends, by predicting the probability of a certain action being the last in the current session, which we denote as the End-of-Session action.
A session consists of a number of actions being either reading or watching some material $M$ or answering a question $Q$, for example:
\begin{align}
    M &\rightarrow 
    M_{\textit{lesson changed}} \rightarrow 
    M_{\textit{lesson changed}}^{\textit{topic changed}} \rightarrow \\ \nonumber
    Q &\rightarrow 
    Q \rightarrow 
    Q_{\textit{lesson changed}} 
    \rightarrow M  
\end{align}
where the sub- and superscript denote if the lesson or topic id was changed compared to the previous action. To each of these actions we associate a binary label indicating whether or not the action is the last action in the session. We consider this binary label a probability of an End-of-Session action, such that the last action has a probability of 100\% for ending the session and all the previous ones have 0\%. Naturally, this leads a to a very imbalanced dataset.
The goal of this task is to create a model able to assign a probability score to each action indicating its End-of-Session probability, and such that it is largest at the true End-of-Session action. 

Usage of the Edulab system is very non-restrictive. A student can engage in any material she wants, and can choose to do the assigned homework at any time. Due to this it is inherently individual how each student uses the system, and when they end their session. It is therefore relevant to consider the problem on a student level, as students have their own preferences with regards to session length, and consequently when the End-of-Session action occurs. This kind of individual behaviour does not only define the general length of student session, but can also influence the dynamics between sessions, since e.g. some students may prefer to alternate between longer and shorter sessions; or take a quick repetition session the next time they engage with the system.

\subsection{Deep Learning Approach \label{our-approach}}
We propose to use a deep learning approach utilizing a recurrent neural network (RNN) for modelling the End-of-Session problem. This approach has the benefit of not requiring complex feature engineering, and is able to capture complex temporal representations of student behaviour. Certain behaviour patterns may occur at a much earlier point than the current action, so to this end we choose to use a Long Short Term Memory (LSTM) network \cite{lstm}.

\begin{figure}
    \centering
    \includegraphics[width=1\linewidth]{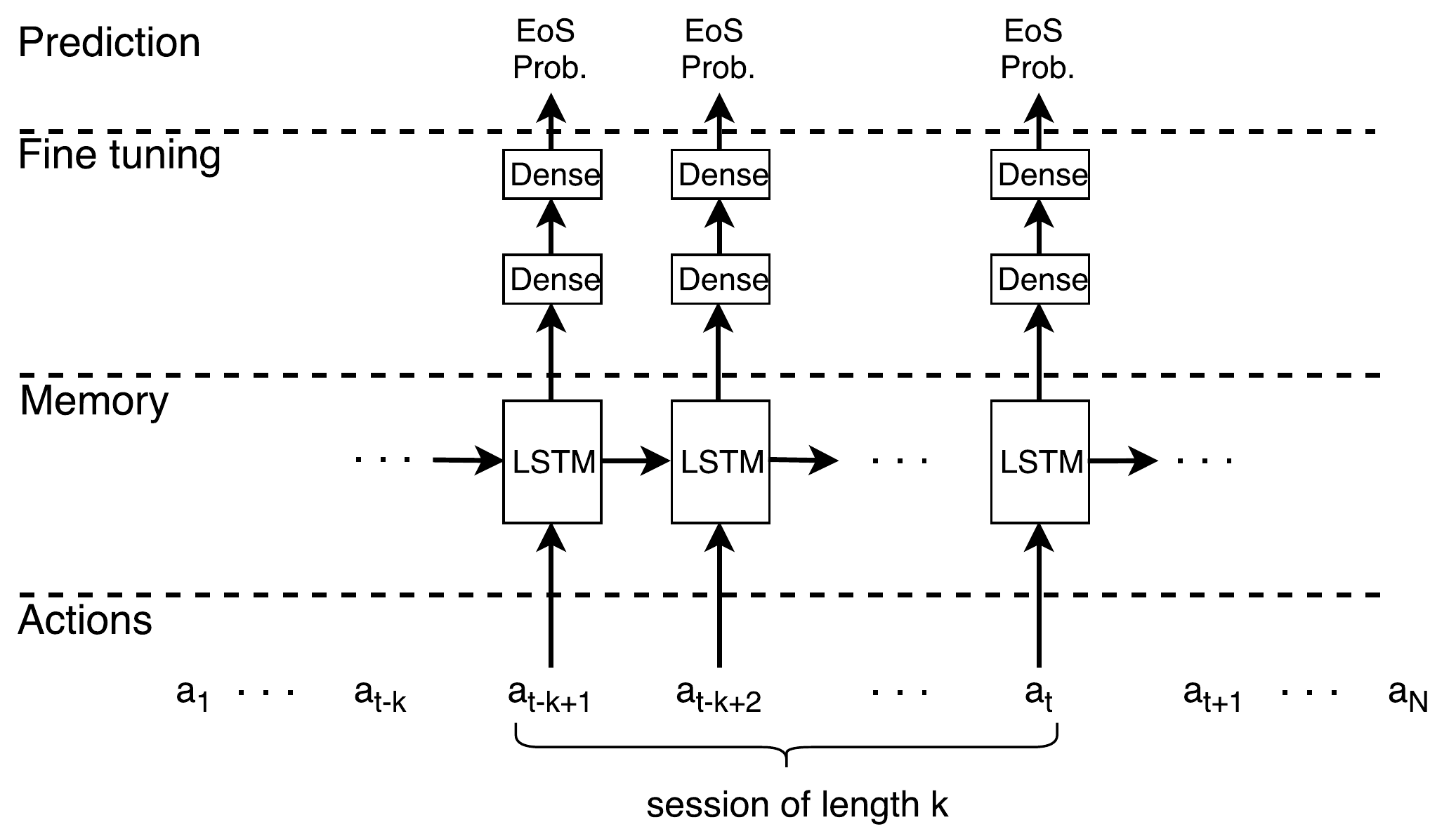}
    \vspace{-10pt}
    \caption{Network architecture of our model.}
    \vspace{-10pt}
    \label{fig:model}
\end{figure}

Our proposed network architecture is displayed in Figure \ref{fig:model}, and can be considered in 4 parts:
\begin{itemize}
    \item \textbf{Actions}: The input to the network are the actions associated with each student. During training each action is labeled either 0 or 1, depending on whether the action is the last in a session. The actions are passed as one long sequence. Each session within this sequence is identifiable by the network through the "time since last session" feature, as well as if the previous label was equal 1, in which case the next action must be a new session.
    \item \textbf{Memory}: The purpose of the memory layer is to let the current interaction's effect on the End-of-Session probability be influenced by previous actions and learned student characteristics. As mentioned previously, we use a LSTM unit for this task, which is able to handle long dependencies in the sequential data, and suffer fewer problems related to vanishing gradients.
    \item \textbf{Fine tuning}: The output of the memory layer is passed to the fine tuning block consisting of two fully connected layers with ReLU activation. The purpose of this is to refine the output of the LSTM, such that the LSTM is able to focus on learning the general underlying behaviour of a student. 
    \item \textbf{Prediction}: The fine tuned output is passed to a single neuron with a sigmoid activation. The sigmoid function returns an output between 0 and 1, representing the End-of-Session probability.
\end{itemize}
Between all layers we use Dropout \cite{dropout}, which sets the activation of a neuron to zero with a certain probability, and is used to limit the network's ability to overfit. 
\subsection{Parameters and training}
For the number of neurons in each layer we use a fan-in approach, where the size is halved from each adjacent layer to the next. Due to the training time of the network, extensive tuning of the dropout probability $p$ and layer sizes is beyond the scope of this paper, but initial experiments showed that $p=0.4$ performed well, which is within the typical range of 0.2 to 0.5 \cite{dropout}. For the layer sizes, initial experiments showed that a LSTM size of 400 performed well, leading to the fine tuning layers having size 200 and 100 respectively.

We train the network using the RMSprop optimizer with a learning rate of 0.001, and use binary cross entropy as the loss function. Due to the End-of-Session problem being heavily class imbalanced, we do a studentwise re-weighting such that if a student has 400 actions consisting of 16 sessions, then each End-of-Session action is weighted by 25, while the others have a weight of 1. This forces the network to balance its ability to predict both normal and End-of-Session actions. This re-weighting changes from student to student based on each student's average session length.
\subsection{Session level model}
In the previous section we presented our approach for modelling the End-of-Session probability by utilizing the full history of past actions for each student. As a baseline approach we can use the same model architecture, but do the modelling on a session level, using just the past actions in the current session, instead of on a student level as before. This will show the benefit of personalizing the model, instead of just focusing on each individual session alone.
\section{Experimental Evaluation}
\label{sec:experimentalpart}
In this section we will describe the experimental evaluation and error analysis of our approach. Our experimental performance comparison is between the performance of the student level model versus the session level model, in order to quantify to what degree a session is self-explanatory with regards to its length, and whether learning a student representation is beneficial. Based on this we analyze the student level model in the following ways:
\begin{enumerate}
    \item We investigate how the End-of-Session probabilities associated with each action increases or decreases throughout a session.
    \item We investigate how homework sessions influence the model's performance, since these can be considered fixed compared to when students use the system on their own initiative.
    \item We consider how the model performs for sessions of varied length, in order to investigate the necessary session length for the model to perform well.
    \item We consider how the model performs on students with limited system usage, i.e. when considering students with a varying number of sessions, in order to investigate how many sessions are needed for this task.
\end{enumerate}
\subsection{Experimental Setup}
We split the students in our dataset randomly into a training set, validation set, and testing set. We use 90\% of the students for training and the remaining 10\% for testing. For validation 10\% of the students from the training data are used. Thus, the students validated and tested on have not been seen previously during training.
For training the network we use a batch size of 64, and employ early stopping with a patients of 3, i.e. we use the model with the best validation performance, and stop after no improvements have been seen in 3 epochs. We use the same approach for training the session level model, where the sessions are extracted from the students in each of the sets.

For measuring the performance we use the area under the receiver operator characteristic curve (AUC). The reason for this is that we do not require the predicted End-of-Session probabilities to be either 0 or 1 (as their labels), but rather expect them to increase slightly over time, with a large increase in the close proximity of the End-of-Session action. Since the sessions are of vastly varied lengths (see Figure \ref{fig:seslenhist}), we argue that this measure is very useful, since it handles the class imbalance, and most importantly provides a measure for how we are able to rank normal and End-of-Session probabilities compared to each other across all students.
\subsection{Model Performance}
\label{sec:model-performance}
In section \ref{our-approach} we presented two models: The first being the student level model where we used all previous actions when predicting the End-of-Session probability of an action, and the second being the session level model where we considered each session alone. We do not expect the session level model to perform as well as the student level model, but this comparison will show to what extent a student's current behaviour can be described by just the current session.
\begin{table}
    \centering
    \begin{tabular}{|l|c|}
         \hline
         Model & AUC \\ \hline
         Student level model & 0.8103 \\ \hline
         Session level model & 0.5647 \\ \hline
    \end{tabular}
    \caption{AUC scores of the student and session level model.}
    \vspace{-10pt}
    \label{tab:modelresults}
\end{table}
Table \ref{tab:modelresults} displays the AUC scores for each of the models. The student level model significantly outperforms the session level model (AUC of 0.8103 vs 0.5647 respectively), showing the importance of modelling the student.

\begin{figure}
    \centering
    \includegraphics[width=0.48\textwidth]{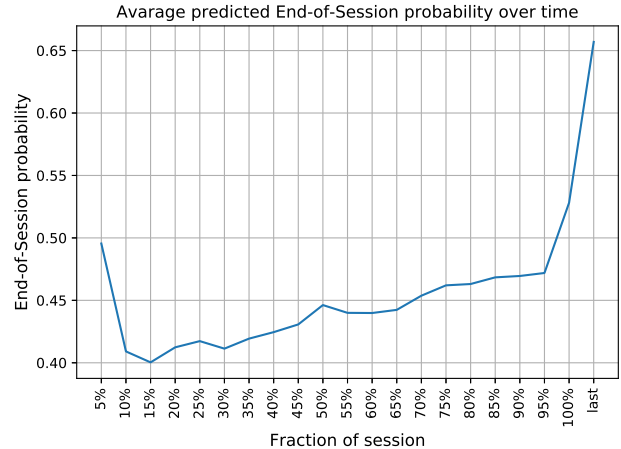}
    \vspace{-20pt}
    \caption{Average End-of-Session probability for sessions split into 5\% chunks, done for sessions with at least 20 actions.}
    \vspace{-10pt}
    \label{fig:eosProbTime}
\end{figure}
We will now consider how the End-of-Session probabilities associated with each action increase or decrease throughout a session. To do this we consider all sessions with at least 20 actions, and cut each session into intervals consisting of 5\% of the actions. This means that each interval for a session with length 20 consists of 1 action, while one of length 100 has 5 actions in each interval. For each of these intervals we compute the average End-of-Session probability. We do this for all sessions and plot the average End-of-Session probabilities for each of the intervals in Figure \ref{fig:eosProbTime}. 

Surprisingly, we observe that the End-of-Session probability is relatively large in the beginning of a session, most notably in the first interval, but also in the second compared to the third. The reason for this is most likely the large amount of short sessions in the dataset (see Figure \ref{fig:seslenhist}), such that the model learns that a student is likely to quit relatively fast. These short sessions could have been removed, but we wanted to base the model on the full history of sessions, and not with artificial session gaps. 

When the model observes that the session has not ended very early, the End-of-Session probability drops, and a small gradual increase is seen until the 95\% interval. In the last interval the End-of-Session probability increases relatively quickly, compared to the gradual increase observed from 10\% to 95\%. The true End-of-Session action has an average probability of 65.7\%. Even though the true End-of-Session action should ideally have a probability closer to 100\%, it is still significantly larger than earlier action's End-of-Session probabilities, which means we are able to create a ranking useful for measuring when a session is likely to end.

\subsection{Further analysis}
We conduct a further analysis of the model focusing on how the performance varies when considering: 1) sessions with varying degree of homework, 2) sessions with varying length, and 3) students with varying number of sessions.
\begin{table}
    \centering
    \begin{tabular}{|l|c|c|}
    \hline
        Sessions consisting of: & AUC \\ \hline
        Only homework & 0.8228 \\ \hline
        Partly homework & 0.8214 \\ \hline
        No homework & 0.7800 \\ \hline
    \end{tabular}
    \caption{The AUC scores of actions in sessions consisting of only homework, partly homework, or no homework actions.}
    \vspace{-10pt}
    \label{tab:homeworkPerformance}
\end{table}
\subsubsection{Homework sessions}
As described in Section \ref{sec:data} students are assigned homework through the Edulab system, where, as seen in Table \ref{tab:dataStats}, 48.3\% of all sessions consist purely of homework. Sessions consisting of partly homework or no homework are at 25.5\% and 26.2\% respectively. We consider how the model performs in each of the three cases, where the AUC scores can be seen in Table \ref{tab:homeworkPerformance}.

Sessions consisting of pure homework can be considered fixed in the sense that a teacher has decided their exact content. Thus, one could imagine the general behaviour of a student to be of lesser importance, since a student is just completing the assigned task. However, if a student is likely to quit early, or split the homework into multiple sessions, then previous observed behaviour is still valuable. The AUC score of all sessions in this case is 0.8228, which is marginally larger than the global AUC score of 0.8103, and thus shows this case to be slightly easier to predict. Sessions consisting partly of homework obtain a similar AUC score of 0.8214. This shows that sessions with fewer individual choices are easier to model (unlike sessions without any constraints).

Sessions consisting of no homework, thus only of actions decided by the student, obtain an AUC score of 0.78, thus lower than the global AUC. This follows our intuition and previous observations, since these should be more difficult, as it is entirely up to the student to define their own task and amount of time they spend using the system.
\subsubsection{Sessions of varying length}
In this section we consider how the model performs on sessions of varying length, where AUC scores across intervals of session lengths can be seen in Table \ref{tab:ses_len_interval}. We see that the model is less accurate when modelling very short sessions of length 1-5 (0.6156 AUC), but from lengths 11 and upwards an AUC of $0.80\pm 0.02$ is obtained. That the model performs worse on very short sessions is to be expected, since a certain amount of actions are needed to infer the student's current behaviour. Additionally, we observed in Section \ref{sec:model-performance} that the model had a tendency to initially predict large End-of-Session probabilities for all actions, which also explains why the model generally performs poorly for very short sessions.

A large part of the Edulab content is presented by groupings of related questions, e.g. as 5 or 10 quick questions, and the number of assigned homework questions are often divisible by 5 as well. Due to this a large part of the sessions have lengths divisible by 5, and the model should be able to detect and adapt to when these kind of patterns occur. In Table \ref{tab:ses_len_interval} we see that the AUC score of those sessions are 0.8944, which is significantly larger than the global AUC, thus showing that the model is indeed able to detect these patterns. In the case of less structured sessions, i.e. those of lengths not divisible by 5, we obtain a significantly lower AUC of 0.7568. These trends are similar to those observed when considering the amount of homework a session consisted of, and shows that while the model is more accurate when structure is present, it is also able to adapt and provide reasonably high AUC scores in the unstructured case.
\begin{table}
    \centering
    \begin{tabular}{|l|c|}
    \hline
         Session length interval & AUC \\ \hline
         1-5 & 0.6156  \\ \hline
         6-10 & 0.7604  \\ \hline
         11-20 & 0.7859  \\ \hline
         21-30 & 0.8200  \\ \hline
         31-40 & 0.8060  \\ \hline
         41-50 & 0.8333  \\ \hline
         51-60 & 0.8109  \\ \hline
         61-70 & 0.8088  \\ \hline
         71-80 & 0.8157  \\ \hline
         81-90 & 0.8103  \\ \hline
         91-max & 0.8016  \\ \hline
         \hline
         Divisible by 5 & 0.8944 \\ \hline
         Not divisible by 5 & 0.7568 \\ \hline
    \end{tabular}
    \caption{The AUC scores of actions in sessions of varying length, grouped in small intervals. E.g. 6-10 corresponds to actions associated with sessions of length 6 to 10.}
    \vspace{-10pt}
    \label{tab:ses_len_interval}
\end{table}
\subsubsection{Students with varying number of sessions}
We consider how the model performs for students with varying system usage, i.e. who has completed a varying number of sessions. In Table \ref{tab:student_ses_counts} we generally see that actions associated with students with more sessions are more accurately predicted, since the intervals of those with more completed sessions obtain larger AUC scores. This shows that the network is able to learn student specific characteristics which are helpful for the task of predicting End-of-Session.
\section{Conclusion}
\label{sec:conclusion}
We presented the problem of determining when a session ends, by modelling the probability of an action being the last in its session, which we denoted as the End-of-Session probability. This problem is difficult to model since we do not get any explicit information about a student's intent to end a session soon, but only in the action it happens. Additionally, a multitude of reasons for the session ending exist, e.g. by a student finishing an assigned homework task, feeling unmotivated, or when the student has mastered the material.
To model this problem we proposed a deep recurrent neural network architecture, that predicts the End-of-Session probability for each student action, by incorporating information from past actions in the current and previous sessions. We consider this a student level model, and for comparison we created a similar session level model, where only the actions associated with a given session were used, and not all previous actions as in the student level model. The student level model obtained an AUC of 0.8103 and the session level model an AUC of 0.5647, thus showing the benefit of learning the student behaviour for this task. Through a detailed error analysis we showed that our model is robust across sessions of different lengths, except for very short sessions of 1 to 5 actions. Similarly, the model gets progressively better for students with more completed sessions, but even with just 1 to 5 sessions an AUC of 0.7799 was obtained. Lastly, the model performed better in sessions with a known structure (e.g. homework), but it was still able to adapt and perform well in sessions where the students chose the content on their own.

In the future we will apply our model, to obtain real-time insights into how the End-of-Session probability progresses throughout sessions and student skill evolution \cite{hansen2014temporal}. Additionally, we will combine this problem with Knowledge Tracing, where relatively simple temporal features have already shown to increase performance \cite{zhangEmbedding}. In terms of modelling, we will explore more elaborate RNN extensions that have been shown to work well with various sequence-based data \cite{hansenSpotifyComp,hansen2019neuraliclr}.

\begin{table}
    \centering
    \begin{tabular}{|l|c|}
    \hline
         Student session interval & AUC \\ \hline
         1-5 & 0.7799  \\ \hline
         6-10 & 0.7866 \\ \hline
         11-20 & 0.7938 \\ \hline
         21-30 & 0.8022 \\ \hline
         31-40 & 0.8054 \\ \hline
         41-50 & 0.8123 \\ \hline
         51-60 & 0.8218 \\ \hline
         61-70 & 0.8236 \\ \hline
         71-80 & 0.8221 \\ \hline
         81-90 & 0.8256 \\ \hline
         91-max & 0.8153 \\ \hline
    \end{tabular}
    \caption{The AUC scores of actions in sessions of students with varying number of total sessions, grouped in small intervals. E.g. 6-10 corresponds to students with between 6 to 10 completed sessions.}
    \vspace{-10pt}
    \label{tab:student_ses_counts}
\end{table}
\section{Acknowledgments}
The work is partly supported by the Innovation Fund Denmark through the DABAI project (5153-00004A).
\bibliographystyle{acm}

\end{document}